\documentclass[twocolumn,aps,floats,showpacs,superscriptaddress,floatfix]{revtex4}
\usepackage[final]{epsfig}
\usepackage{amsmath, amsthm, amssymb}

\newcommand{\be}{\begin{equation}}
\newcommand{\ee}{\end{equation}}
\newcommand{\beas}{\begin{eqnarray*}}
\newcommand{\eeas}{\end{eqnarray*}}
\newcommand{\bea}{\begin{eqnarray}}
\newcommand{\eea}{\end{eqnarray}}

\begin{document} 
\title{Preferential attachment in the growth of social networks: the
case of Wikipedia}
\author{A. Capocci}
\affiliation{Centro Studi e Ricerche E. Fermi, Compendio
  Viminale, Roma, Italy}
\author{V.~D.~P. Servedio}
 \affiliation{Dipartimento di Informatica e Sistemistica, 
Universit\`a di Roma ``La Sapienza", Via Salaria 113, 00198 Roma, Italy}
\affiliation{Centro Studi e Ricerche E. Fermi, Compendio
  Viminale, Roma, Italy}
\author{F. Colaiori}
\affiliation{CNR-INFM(SMC) 
Istituto dei Sistemi Complessi and Dipartimento di Fisica, 
Universit\`a di Roma ``La Sapienza", 
Piazzale Aldo Moro 2, 00185, Roma, Italy}
\author{L.~S.~Buriol}
 \affiliation{Dipartimento di Informatica e Sistemistica, 
Universit\`a di Roma ``La Sapienza", Via Salaria 113, 00198 Roma, Italy}
\author{D. Donato}
 \affiliation{Dipartimento di Informatica e Sistemistica, 
Universit\`a di Roma ``La Sapienza", Via Salaria 113, 00198 Roma, Italy}
\author{S. Leonardi}
 \affiliation{Dipartimento di Informatica e Sistemistica, 
Universit\`a di Roma ``La Sapienza", Via Salaria 113, 00198 Roma, Italy}
\author{G. Caldarelli} 
\affiliation{CNR-INFM(SMC) 
Istituto dei Sistemi Complessi and Dipartimento di Fisica, 
Universit\`a di Roma ``La Sapienza", 
Piazzale Aldo Moro 2, 00185, Roma, Italy}
\affiliation{Centro Studi e Ricerche E. Fermi, Compendio
  Viminale, Roma, Italy}

\date{\today}
\begin{abstract}
We present an analysis of the statistical properties and growth of the
free on-line encyclopedia Wikipedia.  By describing topics by vertices
and hyperlinks between them as edges, we can represent this
encyclopedia as a directed graph.  The topological properties of this
graph are in close analogy with that of the World Wide Web, despite
the very different growth mechanism.  In particular we measure a
scale--invariant distribution of the in-- and out-- degree and we are
able to reproduce these features by means of a simple statistical
model.  As a major consequence, Wikipedia growth can be described by
local rules such as the preferential attachment mechanism, though
users can act globally on the network.
\end{abstract}
\maketitle 

Statistical properties of social networks has become a major research
topic in statistical physics of scale--free networks
\cite{AB02,DM03,PV04}. Collaboration systems are a typical example of
social network, where vertices represent individuals. In the actors'
collaborations case \cite{BA99}, for instance, the edges are drawn
between actors playing together in the same movie. In the case of firm
boards of directors \cite{DYB03,BC04}, the managers are connected if
they sit in the same board. In the scientific co--authorship networks,
\cite{Newman01} an edge is drawn between scientists who co--authored
at least one paper. Other kinds of networks, such as information ones,
are the result of human interaction: the World Wide Web (WWW) is a
well--known example of such, although its peculiarities often put it
outside the social networks category \cite{newman02}.

In this paper, we analyze the graph of Wikipedia \cite{Wikipedia}, a
virtual encyclopedia on line. This topic attracted very much interest
in recent times\cite{holloway05,zlatic06} for its topology. This system
grows constantly as new entries are continuously added by users
through the Internet.  Thanks to the Wiki software \cite{wiki}, any
user can introduce new entries and modify the entries already present.
It is natural to represent this system as a directed graph, where the
vertices correspond to entries and edges to hyperlinks, autonomously
drawn between various entries by contributors.

The main observation is that the Wikipedia graph exhibits a
topological bow--tie--like structure, as does the WWW. Moreover, the
frequency distribution for the number of incoming (in--degree) and
outgoing (out--degree) edges decays as a power--law, while the
in--degrees of connected vertices are not correlated. As these
findings suggest, edges are not drawn toward and from existing topics
uniformly; rather, a large number of incoming and outgoing edges
increases the probability of acquiring new incoming and outgoing edges
respectively. In the literature concerning scale--free networks, this
phenomenon is called ``preferential attachment'' \cite{BA99}, and it
is explained in detail below.

Wikipedia is an intriguing research object from a sociologist's point
of view: pages are published by a number of independent and
heterogeneous individuals in various languages, covering topics they
consider relevant and about which they believe to be competent. Our
dataset encompasses the whole history of the Wikipedia database,
reporting any addition or modification to the encyclopedia. Therefore,
the rather broad information contained in the Wikipedia dataset can be
used to validate existing models for the development of scale--free
networks. In particular, we found here one of the first large--scale
confirmations of the preferential attachment, or
``rich--get--richer'', rule. This result is rather surprising, since
preferential attachment is usually associated to network growth
mechanisms triggered by local events: in the WWW, for instance,
webmasters have control on their own web pages and outgoing
hyperlinks, and cannot modify the rest of the network by adding edges
elsewhere. Instead, by the ``Wiki'' technology a single user can edit
an unlimited number of edges and topics within the Wikipedia network.

\begin{figure}[t!]
\centerline{
\psfig{figure=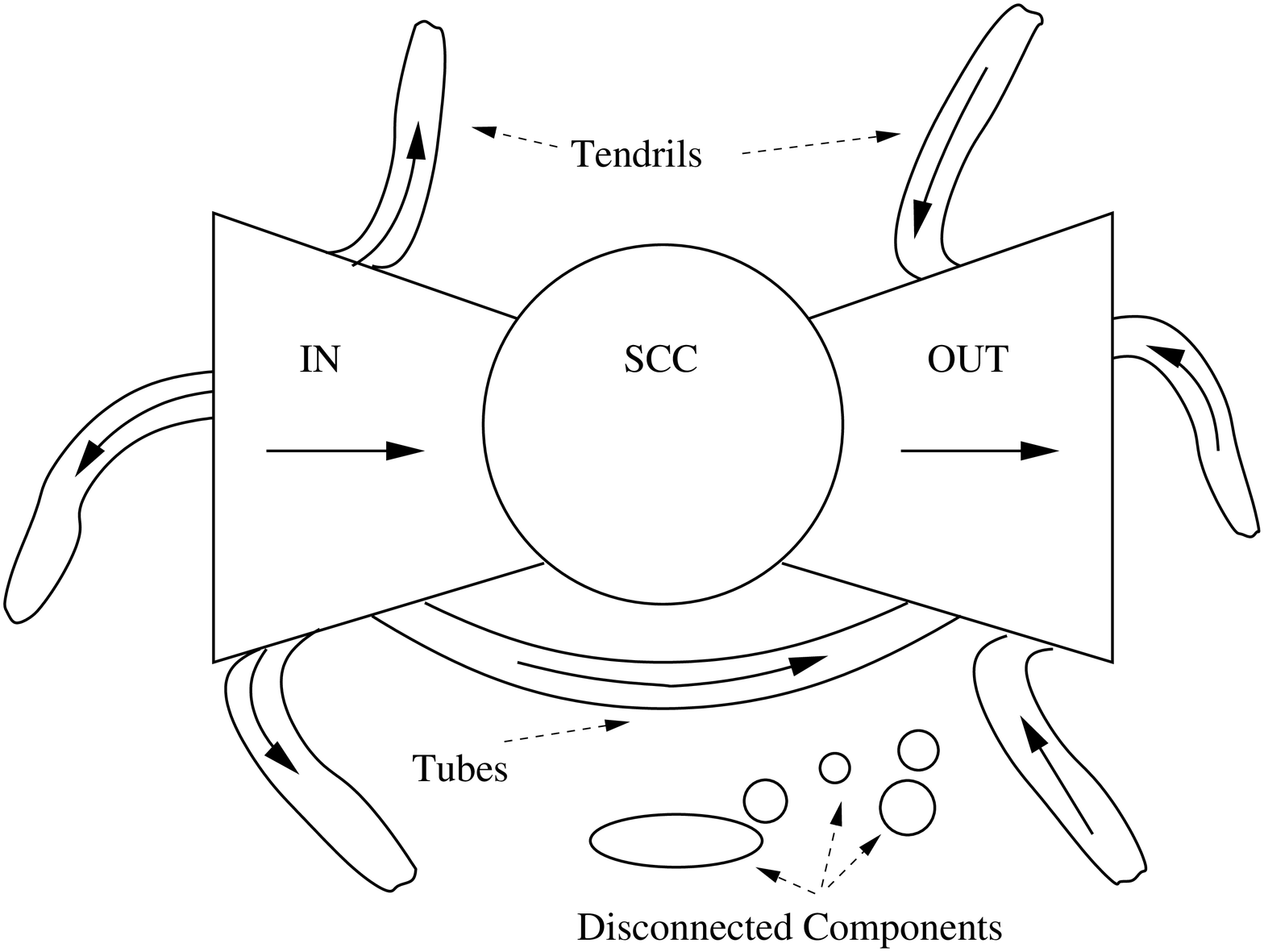, width=7cm}
}
\caption{The shape of the Wikipedia network}.
\label{bowtie}
\end{figure}

The dataset presented here gathers Wikipedia pages in about $100$
different languages; the largest subset at the time of our analysis
was made by the almost $500,000$ pages of the English version, growing
at an exponential pace\cite{voss05}. A detailed analysis of the
algorithms \cite{DLL04b} used to crawl such data is presented
elsewhere \cite{BDLM05}. Here, we start our analysis by considering a
typical taxonomy of regions introduced for the WWW \cite{BKM00}. The
first region includes pages that are mutually reachable by traveling
on the graph, named the strongly connected component (SCC); pages from
which one reaches the SCC form the second region, the IN component,
while the OUT component encompasses the pages reached from the SCC. A
fourth region, named TENDRILS, gathers pages reachable from the IN
component and pointing neither to the SCC nor the OUT region. TENDRILS
also includes those pages that point to the OUT region but do not
belong to any of the other defined regions. Finally TUBES connect
directly IN and OUT regions, and few pages are totally disconnected
(DISC). The result is the so--called bow--tie structure shown in
Fig. \ref{bowtie}.

\begin{table}[h]
\caption{Size of the bow--tie components of the Wikipedia for various
languages. Each entry in the table presents the percentage of vertices
of the corresponding graph that belong to the indicated bow--tie
component.}
\label{t_bowtie-measures}
\begin{tabular}{c|cccccc}
\hline
\texttt{DB}&\texttt{SCC}&\texttt{IN}&\texttt{OUT}&\texttt{TENDRILS}&\texttt{TUBES}&\texttt{DISC} \\
\hline
\texttt{PT}&67.14&6.79&15.85&1.65&0.03&7.50\\
\texttt{IT}&82.76&6.83&6.81&0.52&0.00&3.10\\
\texttt{ES}&71.86&12.01&8.15&2.76&0.07&6.34\\
\texttt{FR}&82.57&6.12&7.89&0.38&0.00&3.04\\
\texttt{DE}&89.05&5.61&3.95&0.10&0.00&1.29\\
\texttt{EN}&82.41&6.63&6.73&0.57&0.02&3.65\\
\hline
\end{tabular}
\end{table}
As a general remark, Wikipedia shows a rather large
interconnection; this means that most of the vertices are in the
SCC. From almost any page it is possible to reach any other. This
feature describes one of the few differences between the on-line
encyclopedia and the WWW: the content of an article can be fully
understood by visiting a connected path along the network.

\begin{figure}[h]
\centerline{
\psfig{figure=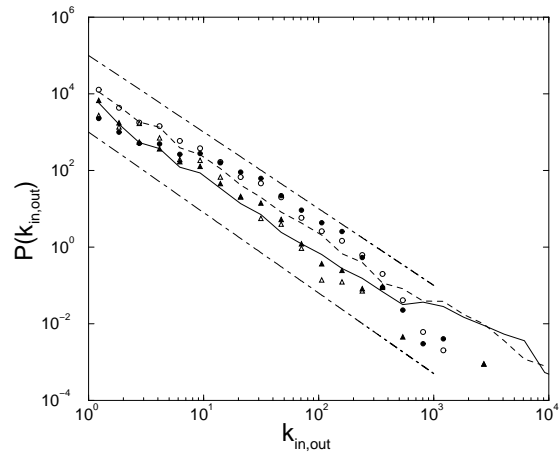, width=7.3cm}
}
\caption{in--degree (white symbols) and out--degree (filled symbols)
  distributions for the Wikipedia English (circles) and Portuguese
  (triangles) graph. Solid line and dashed line represent simulation
  results for the in--degree and the out--degree respectively, for a
  number of 10 edges added to the network per time step. Dot-dashed
  lines show the $k_{in,out}^{-2.1}$ (bottom line) and the
  $k_{in,out}^{-2}$ (top line) behavior, as a guide for the eye.}
\label{f_indegree}
\end{figure}

The key quantities characterizing the structure of an oriented
network are the in--degree ($k_{in}$) and out--degree ($k_{out}$)
distributions. As shown in fig. \ref{f_indegree}, 
both distributions display an algebraic decay, of the kind
$P(k_{in,out}) \propto k_{in,out}^{-\gamma^{in,out}}$, with $2 \leq
\gamma^{in,out} \leq 2.2$.  

Actually, in the case of the out--degree distribution, the value of
the exponent seems to be rather dependent upon the size of the system
as well as the region chosen for the fit. Given the sharp cutoff in
this distribution, the cumulative method of plotting in this case
could result in a quite larger value of the exponent.

We proceeded further by studying the dynamics of the network
growth. The analysis has been made in order to validate the current
paradigm explaining the formation of scale--free networks, introduced
by the Barab\'asi--Albert (BA) model \cite{AB02}. The latter is based
on the interplay of two ingredients: growth and preferential
attachment. In the BA model, new vertices are added to the graph at
discrete time steps and a fixed number $m$ of edges connects each new
vertex to the old ones. The preferential attachment rule corresponds
to assigning a probability $\Pi(k_i) \sim k_{i}$ that a new vertex is
connected to an existing vertex $i$ whose degree is $k_{i}$.  This
elementary process generates a non--oriented network where the degree
follows a power--law distribution.

To observe such a mechanism in a real network, one builds the
histogram of the degree of the vertices acquiring new connections at each
time $t$, weighted by a factor $N(t)/n(k,t)$, where $N(t)$ is the
number of vertices at time $t$ and $n(k,t)$ is the number of vertices with
in--degree $k$ at time $t$.  \cite{Newman01b}.
 
\begin{figure}[h]
\centerline{ \psfig{figure=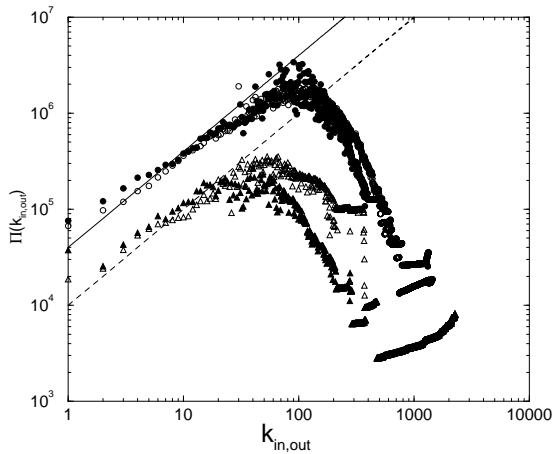, width=7.3cm} }
\caption{The preferential attachment for the in--degree and the
  out--degree in the English and Portuguese Wikipedia network. The
  solid line represents the linear preferential attachment hypothesis
  $\Pi \sim k_{in,out}$.}
\label{prefatc}
\end{figure}

Since the Wikipedia network is oriented, the preferential attachment
must be verified in both directions. In particular, we have observed
how the probability of acquiring a new incoming (outgoing) edge
depends on the present in--(out--)degree of a vertex. The result for the
main Wikipedia network (the English one) is reported in
Fig.\ref{prefatc}.
For a linear preferential attachment, as supposed by the BA model,
both plots should be linear over the entire range of
degrees, here we recover this behavior only partly. 
This is not surprising, since several measurements reported in literature display
strong deviations from a linear behavior \cite{PA05} for large values
of the degree, even in networks with an inherent preferential attachment \cite{Newman01b}. 
It is also possible that for certain datasets (i.e. English), the slope of the 
growth of $\Pi$ is slightly less than $1$.
Nevertheless it is worth to mention that the preferential attachment in Wikipedia  
has a somewhat different nature. Here, most of the times, the edges 
are added between existing vertices differently from the BA model. 
For instance, in the English version of Wikipedia a
largely dominant fraction 0.883 of new edges is created between
two existing pages, while a  smaller fraction of edges points or leaves a
newly added vertex (0.026 and 0.091 respectively).

To draw a more complete picture of the Wikipedia network, we have also
measured the correlations between the in-- and out--degrees of
connected pages. The relevance of this quantity is emphasized by
several examples of complex networks shown to be fully characterized
by their degree distribution and degree--degree correlations
\cite{bianconi05}. A suitable measure for such correlations is the
average degree $K^{(nn)}(k)$ of vertices connected to vertices with
degree $k$ (for simplicity, here we refer to a non--oriented network
to explain the notation). These quantities are particularly
interesting when studying social networks. As other social networks,
collaborative networks studied so far are characterized by assortative
mixing, i.e.\  edges preferably connect vertices with similar degrees
\cite{newman02}. This picture would reflect in a  $K^{(nn)}(k)$
growing with respect to $k$.  If $K^{(nn)}(k)$ (decays) grows with $k$,
vertices with similar degrees are (un)likely to be connected.  This
appears to be a clear cutting method to establish whether a complex
network belongs to the realm of social networks, if other
considerations turn ambiguous \cite{capocci05}.

In the case of an oriented network, such as Wikipedia, one has many
options while performing such assessment: since we could measure the
correlations between the in-- or the out--degrees of neighbor vertices,
along incoming or outgoing edges. We chose to study the average
in--degree $K^{(nn)}_{in}(k_{in})$ of upstream neighbors, i.e. pointing
to vertices with in--degree $k_{in}$. By focusing on the in--degree 
and on the incoming edges, we expect to extract information about the
collective behavior of Wikipedia contributors and filter out their
individual peculiarities: the latter have a strong impact on the
out--degree of a vertex and on the choice of its outgoing edges, since
contributors often focus on a single Wikipedia topic \cite{voss05}.

\begin{figure}[h]
\centerline{
\psfig{figure=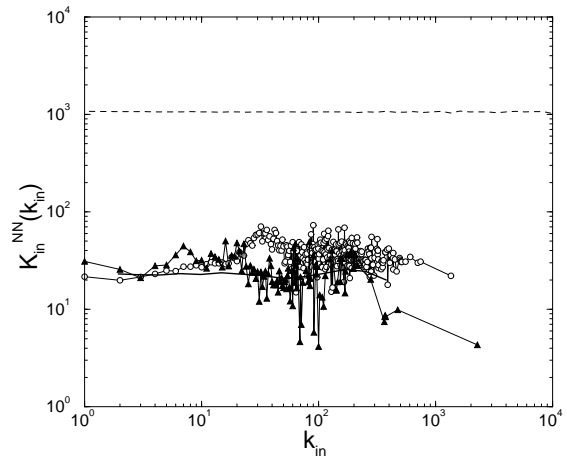, width=7.3cm}
}
\caption{The average neighbors' in--degree, computed along incoming
  edges, as a function of the in--degree for the English (circles) and
  Portuguese (triangles) Wikipedia, compared to the simulations of the
  models for $N=20000$, $M=10$, $R_1=0.026$ and $R_2=0.091$ (dashed
  line) and a realization of the model where the first $0.5\%$ of the
  vertices has been removed to reduce the initial condition impact
  (thick solid line).}
\label{KnninIN}

\end{figure}

Our analysis shows a substantial lack of correlation between the
in--degrees of a vertex and the average in--degree of its upstream
neighboring vertices. So, as reported in fig. \ref{KnninIN}, incoming
edges carry no information about the in--degrees of the connected
vertices, since $K^{(nn)}(k_{in})$ display no clear increasing or
decreasing behavior when plotted against $k_{in}$.

The above quantities, including the power law distribution of the
degrees and the absence of degree--degree correlations, can be modeled
by simple applications of the preferential attachment principle. Let
us consider the following evolution rule, similarly to other models of 
rewiring already considered\cite{krapivsky01}, for a growing
directed network such as Wikipedia: at each time step, a vertex is added
to the network, and is connected to the existing vertices by $M$ oriented
edges; the direction of each edge is drawn at random: with probability
$R_1$ the edge leaves the new vertex pointing to an existing one chosen with
probability proportional to its in--degree; with probability $R_2$,
the edge points to the new vertex, and the source vertex is chosen with
probability proportional to its out--degree. Finally, with probability
$R_3 = 1 - R_1 - R_2$ the edge is added between existing vertices: the
source vertex is chosen with probability proportional to the
out--degree, while the destination vertex is chosen with probability
proportional to the in--degree.

By solving the rate equations for $k_{in}$ and $k_{out}$ by standard
arguments \cite{AB02}, we can show that this mechanism generates power
law distributions of both the in--degree and the out--degree:
$k_{in}$ and $k_{out}$:
\begin{eqnarray}
P(k_{in}) & \simeq & k_{in}^{-\frac{1}{1 - R_2} -1} \nonumber \\
P(k_{out}) & \simeq & k_{out}^{-\frac{1}{1 - R_1} -1}
\end{eqnarray}
which can be easily verified by numerical simulation.

By adopting the values empirically found in the English Wikipedia
$R_1=0.026$, $R_2=0.091$ and $R_3=0.883$, one recovers the same power
law degree distributions of the real network, as shows figure
\ref{f_indegree}.

The degree--degree correlations $K^{(nn)}_{in}(k_{in})$ can be
computed analytically by the same lines of reasoning described in
references \cite{capocci05,barrat05}, and for $1 \ll k_{in} \ll N$ we
have
\begin{equation} \label{theoKNN2}
K^{(nn)}_{in}(k_{in}) \sim \frac{MR_1R_2}{R_3}N^{1 - R_1}
\end{equation}
for $R_3 \neq 0$, the proportionality coefficient depending only
on the initial condition of the network, and 
\begin{equation} \label{theoKNN2b}
K^{(nn)}_{in}(k_{in}) \simeq MR_1R_2\ln N
\end{equation}
for $R_3=0$, where $N$ is the network size. Both equations are
independent from $k_{in}$, as confirmed by the simulation reported in
fig. \ref{KnninIN} for the same values of $R_1$, $R_2$ and $R_3$.

Therefore, the theoretical degree--degree correlation reproduces
qualitatively the observed behavior; to obtain a more accurate
quantitative agreement with data, it is sufficient to tune the initial
conditions appropriately. As shown in fig. \ref{KnninIN}, this can be
done by neglecting a small fraction of initial vertices in the network
model.

In conclusion, the bow--tie structure already observed in the World
Wide Web, and the algebraic decay of the in--degree and out--degree
distribution are observed in the Wikipedia datasets surveyed here. At
a deeper level, the structure of the degree--degree correlation also
resembles that of a network developed by a simple preferential
attachment rule. This has been verified by comparing the Wikipedia
dataset to models displaying no correlation between the neighbors'
degrees.

Thus, the empirical and theoretical evidences show that traditional
models introduced to explain non trivial features of complex networks
by simple algorithms remain qualitatively valid for Wikipedia, whose
technological framework would allow a wider variety of evolutionary
patterns. This reflects on the role played by the preferential
attachment in generating complex networks: such mechanism is
traditionally believed to hold when the dissemination of information
throughout a social network is not efficient and a ``bounded
rationality'' hypothesis \cite{rosvall03,mossa02} is assumed.  In the
WWW, for example, the preferential attachment is the result of the
difficulty for a webmaster to identify optimal sources of information
to refer to, favoring the herding behavior which generates the
``rich--get--richer'' rule.  One would expect the coordination of the
collaborative effort to be more effective in the Wikipedia environment
since any authoritative agent can use his expertise to tune the
linkage from and toward any page in order to optimize information
mining. Nevertheless, empirical evidences show that the statistical
properties of Wikipedia do not differ substantially from those of the
WWW. This suggests two possible scenarios: preferential attachment may
be the consequence of the intrinsic organization of the underlying
knowledge; alternatively, the preferential attachment mechanism
emerges because the Wiki technical capabilities are not fully
exploited by Wikipedia contributors: if this is the case, their focus
on each specific subject puts much more effort in building a single
Wiki entry, with little attention toward the global efficiency of the
organization of information across the whole encyclopedia.
Authors acknowledge support from European Project DELIS.

\end{document}